\documentstyle[11pt]{article}

\newcommand{\K}{\mbox{ , }}
\newcommand{\p}{\mbox{ .}}

\renewcommand{\epsilon}{\varepsilon}

\begin{document}
\title{Kinetic theory for particle production}
\author{Josep Triginer, Winfried Zimdahl\thanks{Present address: 
Fakult\"at f\"ur Physik, Universit\"at Konstanz, PF 5560 M677, 
D-78434 Konstanz, Germany}\  
and Diego Pav\'{o}n\\ 
Departament de F\'{\i}sica\\ 
Universitat Aut\`{o}noma de Barcelona\\ 
08193 Bellaterra (Barcelona), Spain}
\date{}
\maketitle
\thispagestyle{empty}
\begin{abstract}
Recently, the phenomenological 
description of cosmological particle production processes in terms of
effective viscous pressures has attracted some attention. 
Using a simple creation rate model 
we discuss the question to what extent this
approach is compatible with the kinetic theory of a 
relativistic gas. 
We find the effective viscous pressure approach to be consistent with
this model for homogeneous spacetimes but not for inhomogeneous ones.
\end{abstract}
\newpage
\section{Introduction}
It is well known [1 - 3] and widely used [4 - 10] that 
particle production processes in the expanding
Universe may be phenomenologically  described in terms of effective
viscous pressures. 
This is due to the simple circumstance that any source term in the
energy balance of a relativistic fluid may be formally  rewritten in
terms of an effective bulk viscosity. 
The advantage of this rewriting is obvious. 
While an energy momentum balance with a nonzero source term violates
the integrability conditions of Einstein's equations, the energy
momentum balance for a viscous fluid does not. 
Consequently, if it is possible to mimic particle production processes
consistently by effective viscous pressures, one is able to study the
impact of these processes on the cosmological dynamics. 

However, the question remains whether an energy momentum tensor that
formally has been made divergence free by regarding the original
source term as an effective viscous pressure of the cosmic medium, 
adequately describes the physics of the underlying process. 
Is the dynamical effect of a viscous pressure in any respect
equivalent to the corresponding effect of particle production? 
There is no definite answer to these questions at the present state
of knowledge. 

What one would like to have is a microscopic justification of this
effective viscous pressure approach. 
The production processes one has in mind are supposed to play a role
at times of the order of the Planck time or during the reheating
phase of inflationary scenarios. 
It is not unlikely that particle or string 
creation out of the quantum vacuum
has considerably influenced the dynamics of the early Universe
[11, 10, 8]. 
A real microscopic description on the quantum level, however, 
does not seem 
to be available in the near future, and therefore 
it may be useful to study this
phenomenon as far as possible on different classical levels. 
Taking into account that the dynamics of a fluid in or close to 
equilibrium may be derived from kinetic theory, one may ask 
to what extent the phenomenological approach of regarding the
effects of particle production to be in a sense equivalent to those
of viscous pressures may be understood and backed up on the level of
kinetic theory.  
It is  this question that we want to address in the present paper. 
Specifically, we shall model particle production processes by a
nonvanishing source term in a Boltzmann type equation and establish
the link between this kinetic description and the effective viscous
pressure approach on the fluid level. 
It will turn out that the simple phenomenological approach is
compatible with the kinetic theory in homogeneous spaces but not in
inhomogeneous ones. 

One may be reluctant, of course, to assume the applicability of a
kinetic approach in the early Universe. 
A gas, however, is the only system for which the correspondence
between microscopic variables, governed by a distribution function, and
phenomenological fluid quantities is sufficiently well understood. 
All considerations of this paper refer to a model universe for which
the kinetic approach is assumed to be applicable. 
It is the hope that this idealized model nevertheless shares at least
some of the basic features of our real Universe. 

For conventional collisions 
the standard kinetic theory (see, e.g., [12]) is valid if 
the particles interact only weakly, i.e., for systems that are not  
too dense. 
On the other hand, we know from the theory of the strong interaction
that because of the property of `asymptotic freedom' the concept of
weakly interacting particles is especially appropriate for very dense
systems. This might indicate that 
beyond its
conventional range of application, e.g., 
under the conditions of the early Universe, 
the hypothesis of weakly
interacting particles on the level of kinetic theory 
may be not too misleading as well.

In section 2 the basic elements of kinetic theory 
needed in our study are introduced. 
In section 3 our main result is derived in the context of an
`effective rate approximation', modeled after relaxation time
approximations for the Boltzmann collision term. 
In section 4 we summarize our conclusions. 
Units have been chosen so that $c = k_{B} = 1$. 
\section{General kinetic theory}
Our starting point is the assumption that a change in the number of  
particles 
of a relativistic gas should manifest itself in a source term $H$  
on the 
level of kinetic theory.

The corresponding one-particle distribution function $ f =  
f\left(x,p\right)$ 
of a relativistic gas with varying particle number 
is supposed to obey the equation  
\begin{equation}
L\left[f\right] \equiv 
p^{i}f,_{i} - \Gamma^{i}_{kl}p^{k}p^{l}
\frac{\partial f}{\partial
p^{i}}
 = C\left[f\right] + H\left(x, p\right)  \K \label{1}
\end{equation}
where $f\left(x, p\right) p^{k}n_{k}\mbox{d}\Sigma dP$ is the number of
particles whose world lines intersect the hypersurface element 
$n_{k}d\Sigma$ around $x$, having 4-momenta in the range 
$\left(p, p + \mbox{d}p\right)$;  $i$, $k$, $l$ ... = $0$, $1$,  
$2$, $3$. \\
$\mbox{d}P = A(p)\delta \left(p^{i}p_{i} + m^{2}\right) \mbox{d}P_{4}$
is the volume element on the mass shell $p^{i}p_{i} = -  m^{2} $
in the momentum space. 
$A(p) = 2$, if $p^{i}$ is future directed and 
$A(p) = 0$ otherwise; 
$\mbox{d}P_{4} = \sqrt{-g}\mbox{d}p^{1}\mbox{d}p^{2}\mbox{d}p^{3}
\mbox{d}p^{4}$. \\
$C[f]$ is the Boltzmann collision term. 
Its specific structure discussed e.g. by Ehlers [12] will not 
be relevant for our considerations. 
Following Israel and Stewart [13] we shall only require that (i) $C$
is a local function of the distribution function, i.e., independent
of derivatives of $f$, (ii) $C$ is consistent with conversation of
4-momentum and number of particles, and (iii) $C$
yields a nonnegative expression for the entropy production and does
not vanish unless $f$ has the form of a local equilibrium
distribution (see (\ref{9})). 

The term $H(x,p)$ on the r.h.s. of (\ref{1})  takes into account  
the fact
that the number of particles whose world lines intersect a given
hypersurface element within a certain range of momenta may additionally 
change due to creation or decay processes, supposedly of
quantum origin. 
On the level of classical kinetic theory we shall regard this term  
as a given
input quantity. 
Later we shall give an example for the possible functional structure
of $H(x, p)$. 

By the splitting of the r.h.s of eq.(\ref{1}) into $C$ and $H$ we
have separated the collisional from the creation (decay) events. 
In this setting collisions are not accompanied by creation or
annihilation processes. In other words, once created, 
the interactions between the particle 
are both energy-momentum and number preserving. 
Situations like these might be typical, e.g., for the creation of
relativistic particles out of a decaying scalar field during the
reheating phase of standard inflationary universes [8]. 

For a vanishing $H$ eq.(\ref{1})  reduces to the familiar
Boltzmann equation (see, e.g., [12 - 14]). 
In the latter case there exist elaborate solution techniques that
allow to characterize both the equilibrium and nonequilibrium
phenomena, connected with entropy production. 
In the present paper we shall not  study the entropy
production due to collisions within the fluid represented by the term
$C[f]$ on the r.h.s. of (\ref{1}), although the corresponding  
effects may be
additionally included at any stage of our presentation in an obvious
way. 
Instead, we shall focus here on the entropy production due to the  
source term
$H(x, p)$ in equation (\ref{1}).
 
The particle number flow 4-vector 
$N^{i}$ and the energy momentum tensor $T^{ik}$ are
defined in a standard way (see, e.g., [12]) as 
\begin{equation}
N^{i} = \int \mbox{d}Pp^{i}f\left(x,p\right) \K \label{2}
\end{equation}
and
\begin{equation}
T^{ik} = \int \mbox{d}P p^{i}p^{k}f\left(x,p\right) \p \label{3}
\end{equation}
The integrals in (\ref{2}) and (\ref{3}) and throughout this paper
are integrals over the entire mass shell $p^{i}p_{i} = - m^{2}$. 
The entropy flow vector $S^{a}$ is given by [12, 13] 
\begin{equation}
S^{a} = - \int p^{a}\left[
f\ln f - \epsilon^{-1}\left(1 + \epsilon f\right)
\ln\left(1 + \epsilon f\right)\right]\mbox{d}P \K\label{4}
\end{equation}
where bosons and fermions are characterized by $\epsilon = 1$ 
and $\epsilon = -1$, respectively. For $\epsilon = 0$ one recovers
the classical Maxwell-Boltzmann case. 
 
Using the general relationship [15]
\begin{equation}
\left[\int p^{a_{1}}....p^{a_{n}}p^{b}f \mbox{d}P\right]_{;b} 
= \int p^{a_{1}}...p^{a_{n}}L\left[f\right] \mbox{d}P \label{5}
\end{equation}
and eq.(\ref{1}) we find 
\begin{equation}
N^{a}_{;a} = \int \left(C\left[f\right] + H\right) \mbox{d}P \K  
\label{6}
\end{equation}
\begin{equation}
T^{ak}_{\ ;k} =  \int p^{a}\left(C\left[f\right] + H\right) \mbox{d}P
\K \label{7}
\end{equation}
and 
\begin{equation}
S^{a}_{;a} = - \int \ln\left(\frac{f}{1 + \epsilon f}\right)
\left(C\left[f\right] + H\right) \mbox{d}P
\p \label{8}
\end{equation}
In collisional equilibrium, which we
shall assume from
now on, $\ln \left[f/\left(1 + \epsilon f\right)\right]$ in
(\ref{8}) 
is a linear combination of the collision invariants 
$1$ and $p^{a}$.  
The corresponding equilibrium distribution function 
becomes (see, e.g., [12]) 
\begin{equation}
f^{0}\left(x, p\right) = \frac{1}
{\exp{\left[-\alpha - \beta_{a}p^{a}\right]} - \epsilon} 
\K\label{9}
\end{equation}
where $\alpha = \alpha\left(x\right)$ and 
$\beta_{a}$ is a timelike vector that depends on $x$ only.\\
Inserting the equilibrium function into eq.(\ref{1}) one gets
\begin{equation}
\left[p^{a}\alpha_{,a} +
\beta_{\left(a;b\right)}p^{a}p^{b}\right]f^{0} 
\left(1 + \epsilon f^{0}\right)  
=  H\left(x, p\right) 
\p \label{10}
\end{equation} 
\indent It is well known [12, 13] 
that for $H = 0$ this equation, which characterizes 
the `global equilibrium', admits solutions
only for very special cases in which $\alpha = const$ and 
$\beta_{a}$ is a timelike Killing vector. Below we shall discuss the 
corresponding conditions for a specific functional form of $H$ 
in the classical case. 

With (\ref{9}), the balances (\ref{6}) and (\ref{7}) reduce to
\begin{equation}
N^{a}_{;a}=\int H\mbox{d}P,
\label{11}
\end{equation}
and
\begin{equation}
T^{ak}_{;k}=\int p^{a}H\mbox{d}P,
\label{12}
\end{equation}
respectively.

Neither $N^a$ nor $T^{ak}$ are conserved. The nonconservation of  
the number
of particles, the phenomenon under consideration here, is  
accompanied by the
existence of a source term in the energy-momentum balance as well. A 
nonconserved energy-momentum tensor $T^{ak}$, however, is
incompatible with Einstein's field equations. But one may raise the
question whether it is possible to rewrite eq.(\ref{12}) as 
\begin{equation} 
T^{ak}_{;k}-\int p^{a}H\mbox{d}P\equiv\hat{T}^{ak}_{;k}=0,
\label{13}
\end{equation}
with an effective energy-momentum tensor $\hat{T}^{ik}$ that is
conserved and, consequently, is a suitable quantity in Einstein's
equations. Is is basically this problem that we are going to
investigate below.

In collisional equilibrium there is entropy production only due to
the source term $H$. From eq.(\ref{8}) we obtain 
\begin{equation}
S^{a}_{;a} = - \int H\left(x, p\right)
\ln\left(\frac{f^{0}}{1 + \epsilon f^{0}}\right) \mbox{d}P
\K \label{14}
\end{equation} 
implying 
\begin{equation}
S^{a}_{;a} = - \alpha N^{a}_{;a} - \beta_{a}T^{ab}_{\ ;b} \p\label{15}
\end{equation}
Condition (\ref{10}) may be imposed or not. The first possibility
corresponds to the `global equilibrium' case for $H=0$, the second
one to the `local equilibrium' case.
In collisional equilibrium with $f$ replaced by $f^{0}$ in 
(\ref{2}), (\ref{3}) and (\ref{4}), $N^{a}$, $T^{ab}$ and $S^{a}$  
may be 
split with respect to the unique 4-velocity $u^{a}$ according to 
\begin{equation}
N^{a} = nu^{a}\K \label{16}
\end{equation}
and 
\begin{equation}
T^{ab} = \rho u^{a}u^{b} + p h^{ab}\K \label{17}
\end{equation}
with the spatial projection tensor $h^{ab} = g^{ab} + u^{a}u^{b}$ and 
\begin{equation}
S^{a} = nsu^{a}\K \label{18}
\end{equation}
where $n$ is the particle number density, $\rho$ is the energy
density, $p$ is the equilibrium pressure and 
$s$ is the entropy per particle. 
The exact integral expressions for $n$, $\rho$, $p$ and $s$ are given
by the formulae (177) - (180) in [12]. 
According to our general setting described above we assumed 
in establishing (\ref{15}) - (\ref{18}) 
 that
the source term $H$ does not affect the collisional equilibrium. 

Using (\ref{16}) and defining
\begin{equation}
\Gamma  \equiv \frac{1}{n} \int H\left(x, p\right) \mbox{d}P  
\K\label{19}
\end{equation}
eq.(\ref{11}) becomes 
\begin{equation}
\dot{n} + \Theta n = n\Gamma  \p\label{20}
\end{equation}
It is obvious that $\Gamma$ is the particle production rate. 
Similarly, 
with the decomposition (\ref{17}) and the abreviation
\begin{equation}
t^{a} \equiv - \int p^{a}H\left(x, p\right) \mbox{d}P \K\label{21}
\end{equation}
the balance equations for energy and momentum may be written as
\begin{equation}
\dot{\rho} + \Theta \left(\rho + p\right) - u_{a}t^{a} = 0 \K\label{22}
\end{equation}
and
\begin{equation}
\left(\rho + p\right)\dot{u}^{a} + p_{,n}h^{an} + h^{a}_{n}t^{n} = 0
\K \label{23}
\end{equation}
respectively, where $\Theta = u^{a}_{;a}$ is the fluid expansion. 

From the Gibbs equation
\begin{equation}
T\mbox{d}s = \mbox{d}\frac{\rho}{n} + p\mbox{d}\frac{1}{n}\label{24}
\end{equation}
together with (\ref{20}) and (\ref{22}) we find  
\begin{equation}
nT\dot{s} = u_{a}t^{a} - \left(\rho + p\right)\Gamma \p\label{25}
\end{equation}
Inserting (\ref{11}) and (\ref{12}) with  (\ref{19}) and (\ref{21})
into (\ref{15}), the entropy production
density is 
\begin{equation}
S^{m}_{;m} = - \alpha n\Gamma + \beta_{a}t^{a} \p \label{26}
\end{equation}
Using [13]
\begin{equation}
S^{m} = p\beta^{m} - \alpha N^{m} - \beta_{a}T^{ma} \K \label{27}
\end{equation}
which follows 
from (\ref{4}) with (\ref{9}) as well as $\beta_{a} = \beta u_{a}$, and 
(\ref{18}), one finds [12]
\begin{equation}
ns = \beta\left(\rho + p\right)  - \alpha n
\K \label{28}
\end{equation}
and 
(\ref{26}) may be rewritten to yield 
\begin{equation}
S^{m}_{;m} = n\Gamma s + n\dot{s}
\K \label{29}
\end{equation}
where we have used (\ref{25}) and  $\beta \equiv T^{-1}$. 
The latter expression for the entropy production density coincides
with the 
result  that follows from differentiating 
(\ref{18}). 
The first term on the r.h.s. of (\ref{29}) describes the entropy  
production
due to the enlargement of the phase space. The second one takes into
account the contribution due to a possible change in the entropy per 
particle.\\
\indent Henceforth we shall assume, 
that all particles are amenable to a perfect fluid
description immediately after their creation. 
In this case the entropy per particle does not change, i.e., 
$\dot{s} = 0$. 
All particles are created with a fixed entropy.  
Of course, it is always possible to include dissipative effects 
in a standard way [16]. 

According to (\ref{25}) 
the case $\dot{s} = 0$ is equivalent to 
\begin{equation}
u_{a}t^{a} = \left(\rho + p\right)\Gamma \K \label{30}
\end{equation}
relating the source term in the energy balance to that in the
particle number balance. 
The entropy production density 
(\ref{29}) reduces to
\begin{equation}
S^{m}_{;m} = n\Gamma s 
\p \label{31}
\end{equation}
In the perfect fluid case there is entropy production due to the
increase in the number of particles, i.e., due to the enlargement of the
phase space. 

Introducing the quantity
\begin{equation}
\pi \equiv - \frac{u_{a}t^{a}}{\Theta} \K \label{32}
\end{equation}
it is possible to rewrite the energy balance (\ref{22}) as 
\begin{equation}
\dot{\rho} + \Theta \left(\rho + p + \pi\right) = 0 \K\label{33}
\end{equation}
i.e., $\pi$ enters the energy balance in the same way as a bulk viscous
pressure does. 
By (\ref{30}) $\pi$ is related to the particle production rate in the
case $\dot{s} = 0$. 
It had been the hope that a rewriting like this, corresponding to the
introduction of an effectively conserved energy-momentum tensor 
(see (\ref{13})) 
\begin{equation}
\hat{T}^{ik}=\rho u^{i}u^{k}+(p+\pi)h^{ik},
\label{34}
\end{equation}
instead of the nonconserved quantity $T^{ik}$ of (\ref{17}), 
allowed one to
study particle 
production processes in terms of effective viscous pressures. 
Assuming an energy-momentum tensor of the structure (\ref{34}) is
equivalent to map the entire source term in (\ref{12}) onto an
effective bulk viscous pressure. 
It should be mentioned that this is not the only way to take into
account the influence of particle number nonconserving processes on
the cosmological dynamics. Gariel and Le Denmat [17] have developed a
different approach that regards the particle production rate as a
thermodynamic flux on its own. 

The whole procedure leading to (\ref{33}) is rather formal. 
It is unclear, e.g., 
 whether $h^{a}_{n}t^{n}$ in (\ref{23}) will admit a splitting 
\begin{equation}
h^{a}_{n}t^{n} = \pi \dot{u}^{a} + \pi_{,n}h^{an} \K \label{35}
\end{equation}
that would lead to 
\begin{equation}
\left(\rho + p + \pi\right)\dot{u}^{a} + 
\left(p + \pi\right)_{,n}h^{an} = 0
\p \label{36}
\end{equation}
If it were generally true that particle production may be modelled by an
effective bulk viscosity, the splitting (\ref{35}) should be possible.

To clarify this problem, specific expressions for $H\left(x, p\right)$
in (\ref{1}) have to be investigated, while up to this point all
relations are valid for any $H$. 

\section{Effective rate approximation}
\subsection{The basic concept}
As was mentioned earlier, the quantity $H\left(x, p\right)$ is an
input quantity on the level of classical kinetic theory. 
$H$ is supposed to represent the net effect of certain quantum
processes with variable particle numbers (see, e.g., [18]) at the
interface to the classical (nonquantum) level of description. 
Lacking a better understanding of these processes we shall assume
that their influence on the distribution function $f\left(x,
p\right)$ may be approximately 
described by a linear coupling to the latter:
\begin{equation}
H\left(x, p\right) = \zeta\left(x, p\right) f^{0}\left(x, p\right) \p
\label{37}
\end{equation}
Let us further assume that 
$\zeta$ depends on the momenta $p^{a}$ only linearly:
\begin{equation}
\zeta = - \frac{u_{a}p^{a}}{\tau\left(x\right)} + \nu\left(x\right)\p
\label{38}
\end{equation}
$\zeta$, or equivalently $\nu$ and $\tau$, characterize the rate of
change of the distribution function due to the underlying processes
with variable particle numbers. 
The restriction to a linear dependence of $\zeta$ on the momentum  
is equivalent to the requirement that these processes couple to the
particle number flow vector and to the energy momentum tensor 
in the balances for these quantities only,
but not to higher moments of the distribution function.  
This `effective rate approximation' is modeled after the relaxation time
approximations for the Boltzmann collision term [19-21]. 
While the physical situations in both approaches are very 
different, their common feature is the simplified description of 
nonequilibrium phenomena by a linear equation 
for the distribution function in terms of some effective
functions of space and time that characterize the relevant scales of the
process under consideration. 
In the relaxation time approximation this process is determined by
the rate at which the system relaxes to an equilibrium state. 
In the present case the corresponding quantity is the rate by which
the number of particles changes. 

Analogously to the relaxation time model of Maartens and Wolvaardt [21] 
it is possible to find an
exact formal solution of (\ref{1}) with (\ref{37}) 
and (\ref{38}) as well. Equation (1) may be rewritten as
\begin{equation}
L(F)=0.
\label{39}
\end{equation}
The formal solution is
\begin{equation}
F\left(x\left(\eta\right), p\left(\eta\right)\right) 
= f^{0}\left(x\left(\eta\right), p\left(\eta\right)\right)
\exp{\left[-g\left(\eta\right)\right]}\K \label{40}
\end{equation}
with
\begin{equation}
g\left(\eta\right) = \int^{\eta}\mbox{d}\eta ' \zeta
\left(x\left(\eta '\right), p\left(\eta '\right)\right)\K\label{41}  
\end{equation}
$\eta$ being the proper time. 
\subsection{The Maxwell-Boltzmann case corresponding to `global  
equilibrium'}
For a classical gas with $\epsilon = 0$ 
the specific ansatz (\ref{37}) with (\ref{38}) allows us to find  
conditions on
the parameters $\alpha$ and $\beta^{a}$ in (\ref{9}) that replace  
the so called
global equilibrium conditions ($\alpha$=const., $\beta^{a}$-
timelike Killing vector) in the case $H=0$.

Condition (\ref{10}) with $\epsilon = 0$ becomes
\begin{equation}
p^{a}\alpha_{,a}+\beta_{(a;b)}p^{a}p^{b}=\frac{E}{\tau}+\nu,
\label{42}
\end{equation}
where $E=-u_{a}p^{a}$. Decomposing $p^{a}$ according to
$p^{a}=Eu^{a}+\lambda e^{a}$, where $e^{a}$ is a unit spatial vector,
i.e., $e^{a}e_{a}=1$, $u^{a}e_{a}=0$, the mass shell condition
$p^{a}p_{a}=-m^2$ is equivalent to $\lambda^2=E^2-m^2$. The
conditions on $\alpha$ turn out to be
\begin{equation}
\dot{\alpha}=\frac{1}{\tau},\ \ \ h^{a}_{b}\alpha_{,a}=0,
\label{43}
\end{equation}
while $\beta_{a}$ obeys the equation for a conformal Killing vector
\begin{equation}
\beta_{(a;b)}=\Psi(x)g_{ab},
\label{44}
\end{equation}
and
\begin{equation}
m^2\beta_{(a;b)}u^{a}u^{b}=\nu,
\label{45}
\end{equation}
implying
\begin{equation}
\Psi=-\frac{\nu}{m^2}.
\label{46}
\end{equation}
Different from the case $H=0$ where $\alpha$ has to be constant in
space and time, $\alpha$ is only spatially constant in the present
case but changes along the fluid flow lines. From (\ref{45}) 
it follows that 
$\nu=0$ for $m=0$. The condition $\nu=0$ for $m>0$, however, is
equivalent to $\beta_{(a;b})=0$, i.e., the corresponding spacetime is
stationary. Using $\beta_{a}=u_{a}/T$ in (\ref{44}) yields
\begin{equation}
u_{a;b}+u_{b;a}-\left(\frac{T_{,b}}{T}u_{a}+\frac{T_{,a}}{T}u_{b}\right)=
2T\Psi g_{ab}.
\label{47}
\end{equation}
By scalar multiplication with $u^{a}$ one obtains
\begin{equation}
\dot{u}_{b}+\frac{T_{,b}}{T}-\frac{\dot{T}}{T}u_{b}=2T\Psi u_{b},
\label{48}
\end{equation}
and projecting the latter equation in direction of $u^{b}$, the
relation 
\begin{equation}
\frac{\dot{T}}{T}=- \Psi T,
\label{49}
\end{equation}
results [22].

On the other hand, the trace of (\ref{47}) is equivalent to
\begin{equation}
\Theta-\frac{\dot{T}}{T}=4T\Psi.
\label{50}
\end{equation}
Combining (\ref{49}) and (\ref{50}) we find
\begin{equation}
\Theta=-3\frac{\dot{T}}{T}.
\label{51}
\end{equation}
Projecting now (\ref{48}) orthogonally to $u^{a}$ provides us with the
following representation of the acceleration in terms of the spatial 
temperature gradient:
\begin{equation}
\dot{u}_{a}=-(\ln T)_{,b}h^{b}_{a}.
\label{52}
\end{equation}
\indent This relation for a Maxwell-Boltzmann gas 
has an interesting consequence that will be discussed 
at the end of the following section.
\subsection{The balance equations}
Coming back now to the general case that encompasses quantum gases as
well we realize that 
with (\ref{2}), (\ref{37}) and (\ref{38}) 
the particle creation rate (\ref{19}) is given by  
\begin{equation}
n\Gamma  = - \frac{u_{a}}{\tau}N^{a} + Q\left(x\right) 
\label{53}
\end{equation}
for our model, where
\begin{equation}
Q = \nu\left(x\right)M\K\label{54}
\end{equation}
and $M$ is the zeroth moment of the distribution function:
\begin{equation}
M \equiv \int \mbox{d}Pf\left(x, p\right)\p \label{55}
\end{equation}
With (\ref{53}) the particle number balance 
eq.(\ref{20}) may be written in terms of $\tau$ and $\nu$:
\begin{equation}
\dot{n} + n \Theta = \frac{n}{\tau} + Q \p \label{56}
\end{equation}
Similarly, (\ref{2}), (\ref{3}), (\ref{37}), (\ref{38}) and  
(\ref{21}) yield 
\begin{equation}
-t^{a} = - \frac{u_{c}T^{ca}}{\tau\left(x\right)} +  
\nu\left(x\right)N^{a}
\label{57}
\end{equation}
for the source term in the energy momentum balance, 
or, with (\ref{32}), 
\begin{equation}
\Theta \pi  = - \frac{\rho}{\tau} - \nu n 
\p\label{58}
\end{equation}
It is obvious from (\ref{53}) and (\ref{57}) that our effective rate
approximation provides us with a definite coupling between the
functions $\tau$ and $\nu$ that represent the microscopic production
process on a macroscopic level, and the components of the particle
number flow vector $N^{i}$ and the energy momentum tensor $T^{ik}$. 
The expression (\ref{38}) for $\zeta$ ensures that the source terms
(\ref{53}) and (\ref{57}) depend on $M$,  
$N^{i}$ and $T^{ik}$ only, but not on higher moments of the
distribution function. 

 With (\ref{32}) and (\ref{58})  
the energy balance (\ref{22}) becomes 
\begin{equation}
\dot{\rho} + \Theta\left(\rho + p\right) 
= \frac{\rho}{\tau} + \nu n  \p \label{59}
\end{equation}
From 
(\ref{25}) with (\ref{53}) and (\ref{57}) 
it follows that $\dot{s}$ may be 
expressed in terms of $\tau$ and
$\nu$:
\begin{equation}
nT\dot{s} = \nu n - \frac{\rho + p}{n}Q 
- \frac{p}{\tau}  \p \label{60}
\end{equation}
Applying all the steps following (\ref{25}) until (\ref{29}) with
(\ref{53}) for $\Gamma$ and (\ref{57}) for $t^{a}$ we arrive at 
\begin{equation}
S^{m}_{;m} = - \frac{u_{i}S^{i}}{\tau} + Qs + n\dot{s}
\p \label{61}
\end{equation}
By virtue of (\ref{18}) this is equivalent to
(\ref{29}) 
with $\Gamma$ from (\ref{53}). 

Focusing on the perfect fluid case $\dot{s} = 0$ again, we find that 
with (\ref{60}) and (\ref{54}) the general relation (\ref{30}) 
in our specific model reduces to 
\begin{equation}
\nu n = \frac{p}{\tau\left(1 - \frac{\rho + p}{n}\frac{M}{n}\right)}
\p\label{62}
\end{equation}
This equation provides us with a relation between the functions  
$\nu$ and
$\tau$ in the ansatz (\ref{38}) for $\zeta\left(x, p\right)$. 
At this stage it becomes clear why it is necessary to include a function
$\nu$ in (\ref{38}), although intuitively one might have started with 
$\nu = 0$. 
In the latter case the condition $\dot{s} = 0$ seems to require
$p=0$, i.e., our effective rate aproximation seems to apply to a
presureless medium (dust) only if $\tau$ is assumed to be given. But
$p=0$ exactly is an unphysical limiting case (see the integral
expression (179) for $p$ in [12]). 
Consequently,
$\nu=0$ and $\dot{s}=0$ are only compatible
for $\tau^{-1}=0$. There is no particle production at all in this
case.

For a Maxwell-Boltzmann gas ($\epsilon = 0$) a further statement is
possible if eq.(\ref{45}) is imposed. 
A vanishing $m$ in (\ref{45}) neccessarily
implies $\nu=0$.  The obvious
conclusion is that massless particles cannot be produced under this 
requirement. Since the relation (\ref{51}) already holds for massless
particles with $H=0$ [21], this result is hardly surprising. There is no
corresponding restriction, however, in the local equilibrium-like case.

Combining (\ref{53}) and (\ref{54}) with  
(\ref{62}), the production rate  
$\Gamma$ may be written in
terms of $\tau$ as
\begin{equation}
\Gamma = \frac{1}{\tau}\frac{1 - \frac{\rho}{n}\frac{M}{n}}
{1 - \frac{\rho + p}{n} \frac{M}{n}}\p \label{63}
\end{equation}
With (\ref{58}) and (\ref{62}) the effective viscous pressure $\pi$ 
in the energy balance (\ref{33})
is given in terms of $\tau$ as well:
\begin{equation}
\pi = -  \frac{\rho + p}{\Theta\tau}
\frac{1 - \frac{\rho}{n}\frac{M}{n}}
{1 - \frac{\rho + p}{n} \frac{M}{n}}\p \label{64}
\end{equation}
Only for $p<<\rho$ the quantity 
$\Gamma$ approaches $\tau^{-1}$. For $\dot{s} = 0$ the expression 
(\ref{61}) for the entropy production density 
reduces to (\ref{31}) with
$\Gamma$ given by (\ref{63}).

We conclude that as far as the balances  of particle number, energy and
entropy are concerned, the description of particle production in terms
of an effective viscous pressure is backed up by our effective rate
approximation of kinetic theory. 
This is no longer true, however, if the momentum balance (\ref{23})  
comes
into play. 
Eqs. (\ref{16}), (\ref{17}) and (\ref{57}) imply 
\begin{equation}
h^{m}_{a}t^{a} = 0 \p \label{65}
\end{equation}
The momentum balance is completely unaffected by the particle production
rate. 
Consequently, in this respect the particle production is {\it not} 
equivalent to an effective viscous pressure. 
While the simple analogy holds in the homogeneous case, the matter is
more subtle if there are nonvanishing spatial gradients and the fluid
motion is no longer geodesic. 
If one wants to use effective viscous pressures nevertheless to
characterize particle production processes this is consistent only under
the additional assumption that the effective viscous pressure terms
cancel in the momentum balance. 
It is obvious that this condition is equivalent to 
\begin{equation}
\pi\dot{u}^{a} = - \pi_{,n}h^{an}\K \label{66}
\end{equation}
since (\ref{23}) in this case reduces to
\begin{equation}
\left(\rho + p\right)\dot{u}^{a} = - p_{,n}h^{an}\p \label{67}
\end{equation}
Using this equation to eliminate $\dot{u}^{a}$ from (\ref{66}) one gets 
\begin{equation}
\frac{\pi_{,n}}{\pi}h^{an} = \frac{p_{,n}}{\rho + p}h^{an} 
\p \label{68}
\end{equation}
While this condition is empty in a homogeneous spacetime it restricts a
possible spatial dependence of $\pi$.
Especially, it states that a spatially independent $p$ necessarily
implies a spatially independent $\pi$, i.e., a homogeneous creation
rate.  In other words, 
it is impossible to have a homogeneous equilibrium pressure and at the
same time an effective viscous pressure due to particle production that
has a nonvanishing spatial gradient. 

Alternatively, the restriction on $\pi$ may be interpreted
geometrically. 
For a comoving observer the spatial part of the 4-acceleration becomes
\begin{equation}
\dot{u}^{\mu} = \Gamma^{\mu}_{00}\left(u^{0}\right)^{2}\K
\mbox{\ \ \ \ \ \ \ \ \ } (\mu ,\  \nu , ...=\ 1,\ 2,\ 3)\p
\label{69}
\end{equation}
With 
\begin{equation}
u^{0} = \frac{1}{\sqrt{- g_{00}}} \K \label{70}
\end{equation}
and for the rotation free case with 
\begin{equation}
\Gamma^{\mu}_{00} = - \frac{1}{2}g^{\mu\nu}g_{00,\nu} \K \label{71}
\end{equation}
we have, 
\begin{equation}
\dot{u}^{\mu} = \frac{1}{2}g^{\mu\nu}
\left[\ln\left(-g_{00}\right)\right]_{,\nu}
\p \label{72}
\end{equation}
On the other hand, (\ref{66}) is equivalent to 
\begin{equation}
\dot{u}^{\mu} = - g^{\mu\nu}
\left[\ln\left(-\pi\right)\right]_{,\nu}
\p \label{73}
\end{equation}
Consequently, the spatial dependence of $\pi$ is completely determined
by the spatial dependence of $g_{00}$:
\begin{equation}
- \pi = \frac{b\left(t\right)}{\sqrt{-g_{00}}}\K\label{74}
\end{equation}
where $b$ is an arbitrary function of the time. 
Using the relation (\ref{30}) with (\ref{32}), 
$b$ may be expressed in terms of
$\Gamma$:  
\begin{equation}
b\left(t\right) = \frac{\left(\rho + p\right)\Gamma}{\Theta}
\sqrt{-g_{00}}\K \label{75}
\end{equation}
or in terms of $\tau$ if one uses (\ref{63}).

Further relations hold in the Maxwell-Boltzmann 
case ($\epsilon = 0$) corresponding to global
equilibrium. Combining (\ref{52}) and (\ref{67}) we find
\begin{equation}
\frac{T_{,n}}{T}h^{an}=\frac{p_{,n}}{\rho+p}h^{an},
\label{76}
\end{equation}
additionally to (\ref{68}). 
Likewise (\ref{52}) together with (\ref{72}) provides us with
\begin{equation}
(T \sqrt{-g_{00}})_{,\nu}=0
\label{77}
\end{equation}
for a comoving observer. The latter formula for a Maxwell-Boltzmann
gas 
replaces the Tolman relation
$(T\sqrt{-g_{00}})_{,n}=0$ if particles with $m>0$ are produced.
\subsection{Stephani universes and particle production}
\indent Finally, we shall briefly discuss our results concerning the
effective viscous pressure in connection with recent attempts [23]
to find a physical understanding of exact inhomogeneous solutions of
Einstein's field equations with irrotational, shear-free, perfect
fluid sources (Stephani-Barnes family [24, 25]).\\ 
\indent Since there do not
generally exist physically realistic equations of state for the latter
family, Sussman [23] suggested a reinterpretation of these solutions
replacing the perfect fluid source by a fluid with an isotropic bulk
stress. The main advantage of this procedure lies in the introduction
of an additional degree of freedom that might be helpful in finding
reasonable equations of state.

We assume the energy-momentum tensor to be given by 
(\ref{34}) where $\pi$ is to 
describe particle production processes in the manner discussed so
far. While (\ref{34}) was derived for a bounded system with a finite
particle number we shall follow here the common procedure and apply a
quantity like this to describe that part of the early Universe that
developed into its  presently visible part.

In comoving coordinates the metric of the Stephani-Barnes family is 
[22, 25]
\begin{equation}
\mbox{d}s^2=-\frac{(L_{,0}/L)^2}{(\Theta/3)^2}\mbox{d}t^2+L^{-2}(\mbox{d}x^2+ 
\mbox{d}y^2+\mbox{d}z^2),
\label{78}
\end{equation}
\noindent where $\Theta=\Theta(t)$ and $L=L(t,x^{\alpha})$.\\
Generally, for a bulk viscous fluid the balances of energy and
momentum are given by (\ref{33}) and 
(\ref{36}). If $\pi$ mimics the effect of
matter creation then eq.(\ref{36}) reduces to (\ref{67}), 
implying (\ref{68}). With
\begin{equation}
u^{0}=\frac{\Theta}{3}\left (\frac{L_{,0}}{L}\right )^{-1},
\label{79}
\end{equation}
\noindent and 
\begin{equation}
\dot{u}_{\alpha}=\left(\ln\frac{L_{,0}}{L}\right)_{,\alpha},
\label{80}
\end{equation}
\noindent (\ref{33}) and (\ref{67}) may be written as
\begin{equation}
\frac{1}{3}\frac{\rho_{,0}}{L_{,0}/L}=-(\rho +p+\pi),
\label{81}
\end{equation}
\noindent and
\begin{equation}
p_{,\alpha}=-(\rho +p)\left(\ln\frac{L_{,0}}{L}\right)_{,\alpha}
\K\label{82}
\end{equation}
respectively. 
\indent According to the result for our effective rate  
approximation, $\pi$
enters the energy balance but not the momentum balance. At first
sight this seems to be contradictory to Sussman's procedure who
replaced the $p$ of the perfect fluid source by $p+\pi$ in both
balance equations (his equations (3b) and (3c) in [23]). However,
because of (\ref{68}), there exists the additional relation
\begin{equation}
\pi_{,\alpha}=-\pi\left(\ln\frac{L_{,0}}{L}\right)_{,\alpha}.
\label{83}
\end{equation}
Differentiating (\ref{81}) with respect to $x^{\alpha}$ and using 
(\ref{82}) and
(\ref{83}) we find 
\begin{equation}
\frac{\partial}{\partial t}(\rho_{,\alpha}L^3)=0,
\label{84}
\end{equation}
\noindent which is equivalent to
\begin{equation}
L^3\rho_{,\alpha}=f_{,\alpha},
\label{85}
\end{equation}
\noindent where $f=f(x^{\alpha})$ is an arbitrary function. 
Equation (\ref{85}) 
is Sussman's relation (3d) in [23]. Originally derived for a perfect
fluid source, this result remains valid both if $\pi$ is a `real' 
bulk viscous pressure and if it mimics matter creation as in the
present paper.\\
\indent The basic intention of Sussman's reinterpretation is to  
establish a 
`$\gamma$-law' $p=(\gamma-1)\rho$ for the equilibrium pressure $p$
and absorb all terms that do not fit into an equation of state
like this into the quantity $\pi$. With (\ref{81}) the latter is then
{\it defined} by
\begin{equation}
\pi=-\gamma\rho-\frac{1}{3}\frac{\rho_{,0}}{L_{,0}/L}.
\label{86}
\end{equation}
This is again a formal procedure and one has to clarify whether
viscous pressures like this are physically meaningful. Let us
consider as an 
example the subfamily of conformally flat (Petrov type-0)  
solutions, known as 
Stephani universes [24], with
\begin{equation}
L=R^{-1}\{1+\frac{1}{4}k(t)[(x-x_{0}(t))^2+(y-y_{0}(t))^2+(z-z_{0}(t))^2]\},
\label{87}
\end{equation}
\noindent where $R(t)$, $k(t)$, $x_{0}(t)$, $y_{0}(t)$ and  
$z_{0}(t)$ are
arbitrary functions of time. With a perfect fluid source, matter
density and pressure are given by [26]
\begin{equation}
\rho=3C^2(t),
\label{88}
\end{equation}
\noindent and
\begin{equation}
p=-3C^2(t)+2CC_{,0}\frac{L}{L_{,0}},
\label{89}
\end{equation}
\noindent where $C(t)$ is determined by
\begin{equation}
k(t)=[C^2(t)-\frac{1}{9}\Theta^2(t)]R^2(t).
\label{90}
\end{equation}
The Stephani universes are solutions with a homogeneous $\rho$ but
with an inhomogeneous pressure for which, in general, there does not
exist a `$\gamma$-law'. Applying Sussman's procedure to the present
case amounts to replacing $p$ by $p+\pi$ in eq.(\ref{89}):
\begin{equation}
p+\pi=-3C^2+2CC_{,0}\frac{L}{L_{,0}}.
\label{91}
\end{equation}
Since $\rho$ is homogeneous, establishing a `$\gamma$-law' between  
$p$ and 
$\rho$ is equivalent to separate the homogenous part of the r.h.s of
(\ref{91}) and to relate all the inhomogeneities to $\pi$. However,  
this is
clearly inconsistent with (\ref{68}), 
or (\ref{82}) and (\ref{83}) in our specific case.\\
\indent We conclude that a physical interpretation of the Stephani  
universes
with a `$\gamma$-law' ($\gamma=$ constant) and a viscous pressure that
mimics particle production processes is impossible according to our
`effective rate aproximation'.

\section{Concluding remarks}
In the present paper we have investigated the question whether the
widely used description of particle production processes in the early
Universe in terms of effective viscous pressures is compatible with
the kinetic theory of a simple quantum gas. 
Our conclusion is that this approach may be applied, provided the
effective viscous pressure is subject to an additional condition
which entirely fixes its spatial dependence. 
This condition is equivalent to the requirement that all effects due
to particle production cancel in the momentum balance. 
The latter property is a consequence of a simple effective rate
approximation to the source term in the Boltzmann equation that is
supposed to describe particle production. 
While the condition on the spatial behaviour of the effective viscous
pressure $\pi$ is empty in a homogeneous universe, it excludes the
possibility of models with an inhomogeneous $\pi$ as long as the
thermodynamic pressure $p$ remains homogeneous. An explicit example
has been given, showing that the Stephani universes are  
incompatible with a 
bulk viscous pressure associated to matter creation. \\
\  \\
{\bf Acknowledgement} \\
We thank Roberto Sussman and Roy Maartens for useful discussions. 
This paper was supported by  the Spanish 
Ministery of Education under grant 90-676. 
J.T. acknowledges the support of the FPI program.\\ 
\ \\
{\bf REFERENCES} 
\ \\
\begin{small}
\  [1] Zel'dovich Ya B 1970 {\it Sov.Phys.JETP Lett.} {\bf 12} 307\\
\  [2] Murphy G L  1973 {\it Phys.Rev.} {\bf D8} 4231\\
\  [3] Hu B L 1982 {\it Phys.Lett.A} {\bf 90} 375\\
\  [4] Prigogine I Geheniau J Gunzig E and  Nardone P 1989 {\it  
GRG} {\bf 21} 
767\\
\  [5] Calv\~{a}o M O Lima J A S and Waga I 1992 
{\it Phys.Lett A} {\bf 162} 223\\
\  [6] Zimdahl W and Pav\'{o}n D 1993  
{\it Phys.Lett A} {\bf 176} 57\\
\  [7] Zimdahl W Pav\'{o}n D and Jou D 1993 {\it Class. Quantum  
Grav.} {\bf 10} 
1775 \\
\  [8] Zimdahl W and Pav\'{o}n D 1994 {\it Mon. Not. R. Astr. Soc.}  
{\bf 266} 
872\\
\  [9] Zimdahl W and Pav\'{o}n D  1994 {\it GRG} {\bf 26} 1259\\
\ [10] Barrow J D 1988 {\it Nucl.Phys.B} {\bf 310} 743\\
\ [11] Turok N 1988 {\it Phys. Rev. Lett.} {\bf 60} 549\\
\ [12] Ehlers J  1971 in {\it 
General Relativity and Cosmology} ed by Sachs B K 
Academic Press New York\\
\  [13] Israel W and Stewart J M 1979 
{\it Ann Phys} {\bf 118} 341\\
\ [14] de Groot S R van Leeuwen  W A and van Weert Ch G  1980 {\it 
Relativistic Kinetic Theory} North Holland Amsterdam\\
\ [15] Stewart J M 1971 {\it Non-equilibrium Relativistic Kinetic  
Theory} 
Springer New York\\
\ [16] Triginer J and Pav\'{o}n D 1994 {\it GRG} {\bf 26} 513 \\
\  [17] Gariel J and Le Denmat G  1995 
{\it Phys Lett} {\bf A 200} 11\\
\ [18] Birrell D N and Davis P C W 1982 {\it Quantum fields in  
curved space}\\
Cambridge University Press Cambridge \\
\ [19] Marle C 1969 {\it Ann. Inst. H. Poinc\'{a}re} {\bf 10} 67\\
\ [20] Anderson J L and Witting H R 1974 {\it Physica} {\bf 74} 466\\
\ [21] Maartens R and Wolvaardt F P 1994 {\it Class. Quantum Grav.} 
{\bf 11} 203\\
\ [22] Coley A A and Tupper B O J 1990 {\em GRG} {\bf 22} 241\\
\ [23] Sussman R 1994 {\it Class. Quantum Grav.} {\bf 11} 1445\\
\ [24] Stephani H 1967 {\it Commun. Math. Phys.} {\bf 4} 137\\
\ [25] Barnes A 1973 {\it GRG} {\bf 4} 105\\
\ [26] Krasi\'nski A {\it Physics in an inhomogeneous universe} Ch.IV,
preprint, Warsaw 1993\\

\end{small}
\end{document}